\documentclass[prd,a4paper]{revtex4}

\usepackage{graphicx,epsfig,amssymb,subfigure}
\usepackage[latin1]{inputenc}

\font\teneurm=eurm10 \font\seveneurm=eurm7 \font\fiveeurm=eurm5
\newfam\eurmfam \textfont\eurmfam=\teneurm \scriptfont\eurmfam=\seveneurm
\scriptscriptfont\eurmfam=\fiveeurm

\newcommand{\nc}{\newcommand}

\nc{\be}[1]{\begin{equation}\mbox{$\label{#1}$}}
\nc{\bea}[1]{\begin{eqnarray} \mbox{$\label{#1}$}}
\nc{\Section}[2]{\section{#2}\label{#1}}
\nc{\Bibitem}[1]{\bibitem{#1}}
\nc{\Label}[1]{\label{#1}}

\nc{\beq}{\begin{equation}}
\nc{\eeq}{\end{equation}}
\nc{\nom}{\nonumber}
\nc{\bet}{\beta}
\nc{\sq}{\Box}
\nc{\om}{\omega}
\nc{\sig}{\sigma}
\nc{\G}{G}
\nc{\kap}{\kappa}
\nc{\Lam}{\Lambda}

\nc{\eea}{\end{eqnarray}}
\nc{\ee}{\end{equation}}

\nc{\bdm}{\begin{displaymath}}
\nc{\edm}{\end{displaymath}}
\nc{\dpsty}{\displaystyle}
\nc{\bc}{\begin{center}}
\nc{\ec}{\end{center}}
\nc{\ba}{\begin{array}}
\nc{\ea}{\end{array}}
\nc{\bab}{\begin{abstract}}
\nc{\eab}{\end{abstract}}
\nc{\btab}{\begin{tabular}}
\nc{\etab}{\end{tabular}}
\nc{\bit}{\begin{itemize}}
\nc{\eit}{\end{itemize}}
\nc{\ben}{\begin{enumerate}}
\nc{\een}{\end{enumerate}}
\nc{\bfig}{\begin{figure}}
\nc{\efig}{\end{figure}}

\nc{\arreq}{&\!=\!&}
\nc{\arrmi}{&\!-\!&}
\nc{\arrpl}{&\!+\!&}
\nc{\arrap}{&\!\!\!\approx\!\!\!&}
\nc{\non}{\nonumber}
\nc{\align}{\!\!\!\!\!\!\!\!&&}

\def\lsim{\; \raise0.3ex\hbox{$<$\kern-0.75em
      \raise-1.1ex\hbox{$\sim$}}\; }
\def\gsim{\; \raise0.3ex\hbox{$>$\kern-0.75em
      \raise-1.1ex\hbox{$\sim$}}\; }

\nc{\DOT}{\hspace{-0.08in}{\bf .}\hspace{0.1in}}
\nc{\Laada}{\hbox {$\sqcap$ \kern -1em $\sqcup$}}
\nc\loota{{\scriptstyle\sqcap\kern-0.55em\hbox{$\scriptstyle\sqcup$}}}
\nc\Loota{{\sqcap\kern-0.65em\hbox{$\sqcup$}}}
\nc\laada{\Loota}
\nc{\qed}{\hskip 3em \hbox{\BOX} \vskip 2ex}

\nc{\real}{{\rm I \! R}}
\nc{\Z}{{\sf Z \!\!\! Z}}
\nc{\complex}{{\rm C\!\!\! {\sf I}\,\,}}
\def\bigid{\leavevmode\hbox{\small1\kern-3.8pt\normalsize1}}
\def\id{\leavevmode\hbox{\small1\kern-3.3pt\normalsize1}}
\nc{\slask}{\!\!\!/}
\nc{\bis}{{\prime\prime}}
\nc{\pa}{\partial}
\nc{\na}{\nabla}
\nc{\ra}{\rangle}
\nc{\la}{\langle}
\nc{\goto}{\rightarrow}
\nc{\swap}{\leftrightarrow}

\nc{\EE}[1]{ \mbox{$\cdot10^{#1}$} }
\nc{\abs}[1]{\left|#1\right|}
\nc{\at}[2]{\left.#1\right|_{#2}}
\nc{\norm}[1]{\|#1\|}
\nc{\abscut}[2]{\Abs{#1}_{\scriptscriptstyle#2}}
\nc{\vek}[1]{{\rm\bf #1}}
\nc{\integral}[2]{\int\limits_{#1}^{#2}}
\nc{\inv}[1]{\frac{1}{#1}}
\nc{\dd}[2]{{{\partial #1}\over{\partial #2}}}
\nc{\ddd}[2]{{{{\partial}^2 #1}\over{\partial {#2}^2}}}
\nc{\dddd}[3]{{{{\partial}^2 #1}\over
    {\partial #2 \partial #3}}}
\nc{\dder}[2]{{{d #1}\over{d #2}}}
\nc{\ddder}[2]{{{d^2 #1}\over{d {#2}^2}}}
\nc{\dddder}[3]{{d^2 #1}\over
    {d #2 d #3}}
\nc{\dx}[1]{d\,^{#1}x}
\nc{\dy}[1]{d\,^{#1}y}
\nc{\dz}[1]{d\,^{#1}z}
\nc{\dl}[1]{\frac{d\,^{#1}l}{(2\pi)^{#1}}}
\nc{\dk}[1]{\frac{d\,^{#1}k}{(2\pi)^{#1}}}
\nc{\dq}[1]{\frac{d\,^{#1}q}{(2\pi)^{#1}}}

\nc{\bfT}{{\bf T }}

\nc{\cA}{{\cal A}}
\nc{\cB}{{\cal B}}
\nc{\cD}{{\cal D}}
\nc{\cE}{{\cal E}}
\nc{\cG}{{\cal G}}
\nc{\cH}{{\cal H}}
\nc{\cL}{{\cal L}}
\nc{\cO}{{\cal O}}
\nc{\cT}{{\cal T}}
\nc{\cN}{{\cal N}}
\nc{\cR}{{\cal R}}
\nc{\cQ}{{\mathcal{Q}}}
\nc{\rvac}[1]{|{\cal O}#1\rangle}
\nc{\lvac}[1]{\langle{\cal O}#1|}
\nc{\rvacb}[1]{|{\cal O}_\beta #1\rangle}
\nc{\lvacb}[1]{\langle{\cal O}_\beta #1 |}
\nc{\bb}{\bar{\beta}}
\nc{\bt}{\tilde{\beta}}
\nc{\ctH}{\tilde{\cal H}}
\nc{\chH}{\hat{\cal H}}
\nc{\1}{\aa}
\nc{\2}{\"{a}}
\nc{\3}{\"{o}}
\nc{\4}{\AA}
\nc{\5}{\"{A}}
\nc{\6}{\"{O}}
\nc{\al}{\alpha}
\nc{\g}{\gamma}
\nc{\Del}{\Delta}
\nc{\e}{\textrm{e}}
\nc{\eps}{\epsilon}
\nc{\lam}{\lambda}
\nc{\Om}{\Omega}
\nc{\ve}{\varepsilon}
\nc{\mn}{{\mu\nu}}
\nc{\vp}{\varphi}

\nc{\rf}[1]{(\ref{#1})}
\nc{\nn}{\nonumber \\*}
\nc{\bfB}{\bf{B}}
\nc{\bfv}{\bf{v}}
\nc{\bfx}{\bf{x}}
\nc{\bfy}{\bf{y}}
\nc{\vx}{\vec{x}}
\nc{\vy}{\vec{y}}
\nc{\oB}{\overline{B}}
\nc{\oI}{\overline{I}}
\nc{\oR}{\overline{R}}
\nc{\rar}{\rightarrow}
\nc{\ti}{\times}
\nc{\slsh}{\hskip-5pt/}
\nc{\sm}{Standard~Model~}
\nc{\MP}{M_{\rm Pl}}
\nc{\mpl}{M_{\rm Pl}}
\nc{\tp}{t_{\rm Pl}}

\nc{\pmin}{p_{\rm min}}
\nc{\pmax}{p_{\rm max}}
\nc{\fo}{f_0}
\nc{\foi}{f_{0,i}\,}
\nc{\fop}{f_0^P}
\nc{\fou}{f_0^U}

\nc{\eff}{{\rm eff}}
\nc{\MT}{M_{\rm T}}
\nc{\ML}{M_{\rm L}}
\nc{\kk}{\vek{k}}
\nc{\pp}{{\rm p}}
\nc{\pt}{\partial_t}
\nc{\half}{{1\over 2}}
\nc{\w}{\omega}
\nc{\uhat}{\hat{U}_\w}

\nc{\etal}{\mbox{\it et al.}}
\nc{\ie}{{\it i.e. }}
\nc{\eg}{{\it e.g. }}
\nc{\trh}{T_{\rm RH}}
\nc{\ad}{{a'\over a}}
\nc{\bd}{{b'\over b}}
\nc{\Rd}{{R'\over R}}
\nc{\diag}{{\textrm{diag}}}
\nc{\mato}[1]{\tilde{#1}}
\nc{\sinn}{\textrm{sinn}}
\nc{\sech}{\textrm{sech}}
\nc{\I}{\textrm{I}}
\nc{\II}{\textrm{II}}
\nc{\III}{\textrm{III}}
\nc{\vev}[1]{\langle #1 \rangle}
\nc{\hyp}{\,\; F_{1{\hskip -16pt}2}{\hskip 11pt}}
\nc{\brhom}{\overline{\rho}_M}
\nc{\brho}{\overline{\rho}}
\nc{\rhob}{\overline{\rho}}
\nc{\Pb}{\overline{P}}
\nc{\bH}{\overline{H}}
\nc{\ep}{{1+4\eps}}
\nc{\lcdm}{$\Lambda$CDM}
\nc{\wcdm}{$\w$CDM}

\def\smiley{\hbox{\large$\bigcirc$\hspace{-.80em}%
\raise.2ex\hbox{$\cdot\cdot$}\kern-.61em    
\lower.2ex\hbox{\scriptsize$\smile$}}\ }

\def\frowney{\hbox{\large$\bigcirc$\hspace{-.80em}%
\raise.2ex\hbox{$\cdot\cdot$}\kern-.635em
\lower.2ex\hbox{\scriptsize$\frown$}}\ }

\begin{document}

\title{Energy-momentum complexes in $f(R)$ theories of gravity}

\author{Tuomas Multam\"{a}ki} \email{tuomul@utu.fi}
\author{Antti  Putaja}\email{apjput@utu.fi} \affiliation{Department of Physics,
University of Turku, FIN-20014 Turku, FINLAND}
\author{Elias C. Vagenas}
\email{evagenas@academyofathens.gr}
\affiliation{Research Center for Astronomy \& Applied Mathematics\\
 Academy of Athens\\
 Soranou Efessiou 4\\
 GR-11527, Athens, GREECE}
\author{Iiro Vilja}
\email{vilja@utu.fi} \affiliation{Department of Physics,
University of Turku, FIN-20014 Turku, FINLAND}
\date{\today}

\begin{abstract}

Despite the fact that modified theories of gravity, in particular
the $f(R)$ gravity models have attracted much attention in the
last years, the problem of the energy localization in the
framework of these models has not been addressed. In the present
work the concept of energy-momentum complexes is presented in this
context. We generalize the Landau-Lifshitz prescription of calculating the
energy-momentum complex to the framework of $f(R)$ gravity.
As an important special case, we explicitly calculate the energy-momentum complex
for the Schwarzschild-de Sitter metric for a general $f(R)$ theory
as well as for a number of specific, popular choices of $f(R)$.


\end{abstract}
\maketitle

\section{Introduction}

It has been almost a century since the birth of General Relativity
and there are still problems that remain unsolved. The
energy-momentum localization is one of them which till today is
treated as a vexed problem. Much attention has been devoted for this problematic issue.
Einstein was the first who tried to
solve it by introducing the methodology of energy momentum
pseudotensors. He presented the first such prescription
\cite{einstein} and after that a plethora of different
energy-momentum prescriptions were proposed
\cite{ll,tolman,pp,berg,gold,weinberg}. All these prescriptions
were restricted to compute the energy as well as the momenta
distributions in quasi-Cartesian coordinates. M{\o}ller was the
first to present an energy-momentum prescription which could be
utilized in any coordinate system \cite{moller}.

The idea of energy-momentum pseudotensors was gravely criticized for several
reasons \cite{drawbacks1,drawbacks2,drawbacks3,drawbacks4}
(actually one of the drawbacks was the aforesaid use of
quasi-Cartesian coordinates which was solved by M{\o}ller's
prescription).
Firstly, although a symmetric and locally conserved
object, its nature is nontensorial and thus its physical interpretation seemed obscure
\cite{chandra}. Secondly, different energy-momentum complexes could yield different energy distributions
for the same gravitational background \cite{bergqvist1,bergqvist2}. Thirdly, energy-momentum complexes
were local objects while there was commonly believed that the proper energy-momentum of the gravitational
field was only total, i.e. it cannot be localized \cite{chiang}. For a long period of time the idea of
energy-momentum pseudotensors was relinquished.

The approach of energy-momentum pseudotensors
for the thorny problem of energy-momentum localization was
rejuvenated in 1990 by Virbhadra and collaborators
\cite{virbhadra1,virbhadra2,virbhadra3,virbhadra4,
virbhadra5,virbhadra6,virbhadra7,virbhadra8,virbhadra9,virbhadra11}.
Since then, numerous works have been performed on computing the
energy and momenta distributions of different gravitational
backgrounds using several energy-momentum prescriptions (for a
recent list of references see \cite{Vagenas:2006pj}).
In 1996 Aguirregabiria, Chamorro and Virbhadra \cite{virbhadra10} showed that five different
energy-momentum complexes yield the same energy distribution for any Kerr-Schild class metric.
Additionally, their results were identical with the results of Penrose \cite{pen} and Tod \cite{tod}
using the notion of quasi-local mass.
Many attempts since then have been performed to give new definitions of quasilocal energy in
General Relativity \cite{quasilocal1,quasilocal2,quasilocal3,quasilocal4,quasilocal5}. Considerable
efforts  have also been performed in constructing superenergy tensors \cite{senovilla1}. Motivated
by the works of Bel \cite{bel1,bel2,bel3} and independently of Robinson \cite{robinson}, many
investigations have been carried out in this field
\cite{senovilla2, senovilla3,senovilla4,senovilla5,senovilla6}.

In 1999 Chang, Nester and Chen \cite{nester} proved that every energy-momentum complex is associated
with a Hamiltonian boundary term. Therefore, the energy-momentum complexes can be considered as
quasi-local, boundary condition dependent conserved quantities. Finally, it should be pointed out that though
a long way has been trodden the solution to the
problem of energy-momentum localization in the framework of General Relativity is way ahead.

Another challenge to the development of physical theory of gravitation is that,
the plethora of observational data collected recently indicates that our universe is
undergoing an accelerated  expansion. Motivated by this observational evidence, we have been in a
long hunt for the explanation for this speed-up. Till today, three possible reasons have been presented.
Two of them, namely the cosmological constant and the quintessence field, are developed in the framework of
General Relativity. The third one is developed in the framework of alternative theories of gravity.
In particular, the simplest among the aforesaid models are that in which Einstein-Hilbert action is modified
by an additional term.

Modified theories of gravity, especially the $f(R)$ gravity
models that replace the Einstein-Hilbert action of General
Relativity (henceforth abbreviated to GR) with an arbitrary
function of the curvature scalar, have been extensively studied in
recent years (see \eg
\cite{turner,turner2,allemandi,meng,nojiri3,nojiri2,cappo1,woodard,odintsov}
and references therein). The challenges in constructing viable
models in the light of cosmological constraints (see \eg
\cite{new2,Nojiri:2007as,Starobinsky:2007hu} and references
therein), instabilities \cite{dolgov,soussa,faraoni}, solar system
constraints (see {\it e.g.} \cite{chiba,confprobs,Clifton,Hu2007}
and references therein) and evolution large scale perturbations
\cite{Bean:2006up,Song:2006ej,Song2} are now known. The Solar
System constraints are a major obstacle to most theories
\cite{Erickcek:2006vf,Chiba2,Jin:2006if, Faulkner:2006ub} but they
can be completely removed by certain types of models
\cite{Nojiri:2007as, Hu2007, kimmo, Nojiri:2007cq,Cognola:2007zu}.

It is widely known that when a new theory is introduced, it is expected this new theory
to successfully answer all already-solved (in the framework of the old theory) problems.
Moreover, it is anticipated that this new theory will be able
to address, alleviate, and finally solve problems that the existing old theory cannot.
Following this line of thought, we address here, to our knowledge, for the very first time
the problematic issue of energy-momentum localization in the
context of $f(R)$ gravity models which as was mentioned above intend to replace GR.
We take first steps in this direction and consider energy-momentum
complexes within $f(R)$ gravity models.

The remainder of the paper is as follows. In Section II, we present
the basic equations and formalism of $f(R)$ gravity. In Sections III and IV,
the Landau-Lifshitz energy-momentum complex and the Schwarzschild-de Sitter (henceforth abbreviated
as SdS) metric, or the SdS black hole background, are reviewed.
In Section V, we extend the
concept of the Landau-Lifshitz energy-momentum complex into the framework of $f(R)$ theories.
As a special case, we compute the energy-momentum complex of the SdS metric
for a number of commonly considered $f(R)$ theories. In the final section we
summarize the results and present our conclusions.

\section{$f(R)$ gravity formalism}

The action for $f(R)$ gravity is
\be{action}
S = \int{d^4x\,\sqrt{-g}\Big(\frac{1}{16\pi G}f(R)+{\cal{L}}_{m}\Big)},
\ee
where the standard Einstein-Hilbert action is replaced by a general function of
scalar curvature $f(R)$.
The corresponding field equations (in the metric approach) are found by varying
with respect to the metric $g_{\mu\nu}$ and read as
\be{eequs}
F(R) R_{\mu\nu}-\frac 12 f(R) g_{\mu\nu}-\nabla_\mu\nabla_\nu F(R)+
g_{\mu\nu}\Box F(R)= 8 \pi G T^m_{\mu\nu}
\ee
where $T_{\mu\nu}^m$ is the standard minimally coupled stress-energy tensor and
$F(R)\equiv df/dR$. In contrast to the standard Einstein's equations from the
Einstein-Hilbert action, the field equations are now
of higher order in derivatives.

Contracting the field equations gives
\be{contra}
F(R)R-2 f(R)+3\Box F(R)=8 \pi G (\rho-3p)
\ee
where we have assumed that we can describe the stress-energy tensor with a
perfect fluid. From the contracted equation it is clear that in vacuum, any
constant scalar curvature metric with $R=R_0$ is a solution of the contracted
equation as long as $F(R_0)R_0=2f(R_0)$. In general, the whole set of field
equations is solved exactly by the SdS
metric \cite{Cognola:2005de}
(for a more recent work on spherically symmetric solutions of modified field
equations in $f(R)$ gravity see also \cite{Multamaki2})
\be{SdS1}
ds^{2}=B dt^{2}-B^{-1}dr^{2}-r^{2}d\theta^{2}-r^{2}\sin^{\!2}\!\theta d\phi^{2}
\ee
with
\be{B}
B(r)=1-\frac {2M}{r}-\Lambda \, \frac {r^2}{3}.
\ee
The scalar curvature for this metric is $R_0=-4\Lambda$.
Hence any $f(R)$ theory, including the standard General Relativity,
satisfying the constant curvature condition
$F(R_0)R_0=2f(R_0)$ has the SdS (black hole) metric as an exact solution.
We will return to this important special case in a later section.

\section{The Landau-Lifshitz Energy-Momentum Complex}

In general, the energy-momentum complex $\tau^{\mu\nu}$ (henceforth abbreviated as EMC) carries
coordinate dependent information on the energy content of the gravitational and
matter fields. It sums up the energies of the matter fields through the
energy-momentum tensor $T^{\mu\nu}$ (henceforth abbreviated as EM), and that of the
gravitational field through the energy-momentum pseudotensor
 $t^{\mu\nu}$ (henceforth abbreviated as EMPT), which
depends on the coordinate system used to describe the system. The EMPT cannot be
defined uniquely and a number of suggestions, with different mathematical properties,
exist. All of them lead to conserved quantities of the gravitational theory.

The most straightforward conserved quantity is the integrated EMC over the
 three-dimensional space integral
\be{intEMC}
E_{EMC} = \int_{B(0,r)} d^3x\, \tau^{00} \label{intEMC1}
\ee
which represents both the energy of the gravitational field and that of matter inside
 the coordinate volume $B(0,r)$. In the case of a black hole, it consists of two parts:
the black hole mass M and the energy stored in the gravitational field $t^{00}$,
therefore
\be{intEMCbh}
E_{EMC} = M + \int_{B(0,r)} d^4x\sqrt{-g}\, t^{00}\label{intEMCbh}.
\ee


In the construction of Landau and Lifshitz \cite{ll}, one looks for an object,
$\eta^{\mn\al}$, that is antisymmetric in its indices since then
$\pa_{\nu}\pa_{\al}\eta^{\mn\al}=0$ due to the covariant continuity equation,
which in a locally Minkowskian coordinate system, simplifies to
\begin{eqnarray}
    \pa_{\nu}T^{\mn}=0.
\end{eqnarray}

 Hence
\begin{eqnarray}
    T^{\mn}=\pa_{\al}\eta^{\mn\al}.
\end{eqnarray}
Einstein's equations give
\begin{eqnarray}
\label{materianeit}
    T^{\mn}=\frac{1}{\kappa^{2}}(R^{\mn}-\frac{1}{2}R g^{\mn})
\end{eqnarray}
where $\kappa ^{2}= 8 \pi G$ (we have set $c=1$), and in the locally
Minkowskian coordinates Ricci tensor and Ricci scalar reads as
\begin{eqnarray}
\label{pmruv}
    R^{\mn}& = & \frac{1}{2}g^{\mu\al}g^{\nu\beta}g^{\gamma\delta}(\pa_{\al}\pa_{\delta}
g_{\gamma\beta}+\pa_{\gamma}\pa_{\beta}g_{\al\delta}\nom\\
    & & -\pa_{\al}\pa_{\beta}g_{\gamma\delta}-\pa_{\gamma}\pa_{\delta}g_{\al\beta})\\
    \mbox{and}\hspace{3ex}R & = & \pa^{\al}\pa^{\bet}g_{\al\bet}-g^{\al\bet}\sq
g_{\al\bet}\nom\\
        & = & (g^{\al\mu}g^{\bet\nu}-g^{\al\bet}g^{\mu\nu})\pa_{\mu}\pa_{\nu}g_{\al\bet}.
\eea
Using these expressions in Eq. (\ref{materianeit}), one can rewrite the EM tensor
as \cite{ll}
\be{tmunu}
T^{\mn}=\pa_{\al}\eta^{\mn\al}
\ee
where
\begin{eqnarray}
\label{etaeinsteinLL}
    \eta^{\mn\al}=\frac{1}{2\kappa^{2}}\frac{1}{(-g)}\frac{\pa}{\pa x^{\beta}}
\left[(-g)(g^{\mn}g^{\al\beta}-g^{\mu\al}g^{\nu\beta})\right].
\end{eqnarray}
In a locally Minkowskian coordinate system  $\pa_{\al}g_{\mn}=0$ and hence one can
define
\begin{eqnarray}
\label{konti}
    (-g)T^{\mn}\equiv\pa_{\al} h^{\mn\al}\equiv\pa_{\al}\pa_{\bet}H^{\mu\nu\al\bet}
\end{eqnarray}
where two new tensors, so-called superpotentials,
\bea{hoo}
\label{konti1}
 h^{\mn\al} & = & (-g)\eta^{\mn\al}=\frac{1}{2\kappa^{2}}\frac{\pa}{\pa x^{\beta}}
\left[(-g)(g^{\mn}g^{\al\beta}-g^{\mu\al}g^{\nu\beta})\right]\\
\label{konti2}
 H^{\mu\nu\al\bet} & = & \frac{1}{2\kappa^{2}}[(-g)(g^{\mn}g^{\al\beta}-g^{\mu\al}
g^{\nu\beta})]
\eea
have been defined.

In a general coordinate system, Eq. (\ref{konti}) is no longer valid and one defines
a new object $t^{\mn}$ such that
\begin{eqnarray}
\label{kokokontiantti}
    (-g)(T^{\mn}+t^{\mn})\equiv\frac{\pa h^{\mn\al}}{\pa x^{\al}}.
\end{eqnarray}
The new object, namely the EMPT $t^{\mn}$, is straightforwardly computed
in a general coordinate system employing Eq. (\ref{kokokontiantti}) since $T^{\mn}$
can be expressed in terms of the geometric quantities by using the Einstein's
equation, {\it i.e.} Eq.
(\ref{materianeit}), and $h^{\mn\al}$ is given in Eq. (\ref{konti1}).

Carrying out this somewhat lengthy but routine exercise, one obtains \cite{ll}
\bea{no3}
t_{LL}^{\mn}=&\frac{1}{2\kappa^{2}}\left\{(2\Gamma^{\eps}_{\al\beta}
\Gamma^{\om}_{\eps\om}-
\Gamma^{\eps}_{\al\om}\Gamma^{\om}_{\beta\eps}-\Gamma^{\eps}_{\al\eps}
\Gamma^{\om}_{\beta\om})(g^{\mu\al}g^{\nu\beta}-g^{\mu\nu}g^{\al\beta})\right.\nom\\
&\left.+g^{\mu\al}g^{\beta\eps}(\Gamma^{\nu}_{\al\om}\Gamma^{\om}_{\beta\eps}+
\Gamma^{\nu}_{\beta\eps}\Gamma^{\om}_{\al\om}-\Gamma^{\nu}_{\eps\om}
\Gamma^{\om}_{\al\bet}-
\Gamma^{\nu}_{\al\beta}\Gamma^{\om}_{\eps\om})\right.\nom\\
&\left.+g^{\nu\al}g^{\beta\eps}(\Gamma^{\mu}_{\al\om}\Gamma^{\om}_{\beta\eps}+
\Gamma^{\mu}_{\beta\eps}\Gamma^{\om}_{\al\om}-\Gamma^{\mu}_{\eps\om}
\Gamma^{\om}_{\al\beta}-
\Gamma^{\mu}_{\al\beta}\Gamma^{\om}_{\eps\om})\right.\nom\\
&\left.+g^{\al\beta}g^{\eps\om}(\Gamma^{\mu}_{\al\eps}\Gamma^{\nu}_{\beta\om}-
\Gamma^{\mu}_{\al\beta}\Gamma^{\nu}_{\eps\om})\right\}.
\eea
The Landau-Lifshitz EMC, {\it i.e.} $\tau_{LL}^\mn$, can now be evaluated either as a sum of
the EM and the EMPT, namely
\be{elias}
\tau_{LL}^\mn=(-g)\left(t_{LL}^{\mn}+T^{\mn}\right)
\ee
or directly using Eq. (\ref{kokokontiantti}) which is now written as
\begin{eqnarray}
\label{kokokontiantti1}
    \tau_{LL}^\mn\equiv\frac{\pa h^{\mn\al}}{\pa x^{\al}}.
\end{eqnarray}

\section{The Schwarzschild-de Sitter metric}

In GR, the empty space solution outside a static spherically symmetric mass
distribution in a universe with a cosmological constant is the
SdS metric, or the SdS black hole metric. In spherically symmetric coordinates, it reads outside the
mass distribution as (in units where $G=1$) given in eqs. (\ref{SdS1}) and (\ref{B}).
$M$ is the total mass and $\Lambda$ is the cosmological constant. Due to the fact that
some EMPTs are calculated in
cartesian coordinates, we need to re-express the SdS black hole metric in cartesian
terms. The metric (\ref{SdS1}) then reads as
\bea{sdeskaart}
ds^{2} & = &B dt^{2}-\frac{B^{-1}x^{2}+y^{2}+z^{2}}{x^{2}+y^{2}+z^{2}}dx^{2}-
\frac{x^{2}+B^{-1}y^{2}+z^{2}}{x^{2}+y^{2}+z^{2}}dy^{2} \nom\\
&- & \frac{x^{2}+y^{2}+B^{-1}z^{2}}{x^{2}+y^{2}+z^{2}}dz^{2}-
\frac{(B^{-1}-1)2yz}{x^{2}+y^{2}+z^{2}}dy dz \nom\\
&- & \frac{(B^{-1}-1)2xy}{x^{2}+y^{2}+z^{2}}dx dy-
\frac{(B^{-1}-1)2xz}{x^{2}+y^{2}+z^{2}}dx dz.
\eea
For this metric, but not for the metric in
spherically symmetric coordinates, the determinant reads as $g=-1$.
This feature is used regularly in
the following as outer factors like $\sqrt{-g}\;$ equal unity.


The EMPTs corresponding to the SdS solution are
straightforwardly calculable (in a cartesian coordinate system).
Working in a space where $r > 0$ and thus EMPTs and EMCs coincide,
we obtain for the 00-component of the  Landau-Lifshitz EMPT
\bea{no14}
\label{sdsll}
    (-g) t^{00} & = & -\frac{2}{\kap^2} \frac{36M^2+12M\Lam r^3+
\Lam r^4(\Lam r^2-9)}{r^2(6M+r(\Lam r^2-3))^{2}}\nom\\
   & = & -\frac{2}{\kap^2} \frac{-9\Lam r^4+(6M+\Lam r^3)^2}{r^2(\Lam r^3-3r+6M)^{2}}.
\eea

For comparison, one can easily repeat this exercise for the other well-know EMPTs: we find that
the Weinberg EMPT is equal to that of Landau and Lifshitz,
the Einstein and Tolman EMPT is
\beq
    t^{0}_{0}=-\frac{\Lam}{\kap^2},
\eeq
and the M\o ller EMPT is
\beq
    t^{0}_{0}=-\frac{2\Lam}{\kap^2}.
\eeq

It is evident that the EMPT of the SdS metric strongly depends on the
chosen construction, even
though EMPTs derived in different prescriptions sometimes can be identical.
Furthermore, their behavior far from the mass source is wildly different.
Therefore, by looking only the functional form of a single EMC, it is obscure what
is the precise physical interpretation of it, even though it
represents a conserved quantity. However, we emphasize, that the different EMCs
differ by a divergence term related to the boundary conditions of the physical
situation \cite{nester}.

\section{EMPT in $f(R)$ theories of gravity}\label{freipt}

Among the different EMPTs studied in the literature, Landau-Lifshitz's and Weinberg's prescriptions
appear to be most straightforwardly suitable for extending into $f(R)$ gravity
theories. Here we consider extending the Landau-Lifshitz's prescription and leave
others for future work.

In $f(R)$ theories, one can write the field equations as
\begin{widetext}
\begin{eqnarray}\label{freit}
T_{\mn}=\frac{1}{\kappa^{2}}\left\{f'(R)R_{\mn}-\frac{1}{2}g_{\mn}f(R)-
D_{\mu}D_{\nu}f'(R)+
g_{\mu\nu}\sq f'(R)\right\}.
\end{eqnarray}
\end{widetext}
Like in GR, the covariant continuity equation holds, i.e.
$D_\mu T^{\mn}=0$ \cite{koivisto},
suggesting that one should write the RHS of Eq. (\ref{freit}) as a divergence of an
object antisymmetric in its indices,{\it i.e.} in a form $\partial_\al h^{\mn\al}$.

Following Landau's and Lifshitz's prescription and considering a locally Minkowskian
coordinate system at a given point we obtain
\begin{widetext}
\bea{freiptlasku}
\kappa^{2}T^{\mn} & = & f'(R)R^{\mn}-\frac{1}{2}g^{\mn}f(R)+(g^{\mn}g^{\al\bet}-
g^{\mu\al}g^{\nu\bet})
\pa_{\al}\pa_{\bet}f'(R)\nom\\
& = & f'(R)G^{\mn}
+\frac{1}{2}g^{\mn}(f'(R)R-f(R))+
\pa_{\al}\left[(g^{\mn}g^{\al\bet}-g^{\mu\al}g^{\nu\bet})\pa_{\bet}f'(R)\right]\nom\\
& = & \pa_{\al}\left[f'(R)\kap^{2}\eta^{\mn\al}+(g^{\mn}g^{\al\bet}-g^{\mu\al}
g^{\nu\bet})f''(R)
\pa_{\bet}R\right]-\pa_{\al}f'(R)\kap^{2}
\eta^{\mn\al}
+ \frac{1}{2}g^{\mn}(f'(R)R-f(R))\nom\\
& = & \pa_{\al}\left[f'(R)\kap^{2}\eta^{\mn\al}+(g^{\mn}g^{\al\bet}-g^{\mu\al}
g^{\nu\bet})f''(R) \pa_{\bet}R\right]+\frac{1}{2}g^{\mn}(f'(R)R-f(R))
\eea
\end{widetext}
where $\eta^{\mn\al}$ is that defined in (\ref{etaeinsteinLL}).
It is noteworthy that the term $\pa_{\al}f'(R)\kap^{2}
\eta^{\mn\al}$ vanishes in locally Minkowskian coordinate system, because
$\eta^{\mn\al}$ is linear in
the first derivatives of the metrics.
We are partially able to
write the RHS of the field equations, namely Eq. (\ref{freiptlasku}), as a
divergence. The remaining term, absent in GR, remains problematic in a general case
without a clear method which would enable us to write it as a four divergence.
We can, however, proceed in an important special case where the scalar curvature is a
constant, $R=R_0$. The SdS black hole metric belongs to such a class of metrics.

\subsection{The Landau-Lifshitz -energy momentum complex for a metric with constant scalar curvature}

For a metric with a constant scalar curvature, $R=R_0$, Eq. (\ref{freiptlasku}) simplifies to
\bea{pseudofr}
T^{\mn} & = & \pa_{\al}[f'(R_{0})\eta^{\mn\al}]+\frac{1}{2\kappa^{2}}g^{\mn}(f'(R_{0})
R_{0}- f(R_{0}))\nom\\
& = & \pa_{\al}[f'(R_{0})\eta^{\mn\al}]+\frac{1}{6\kappa^{2}}\pa_{\al}(g^{\mn}x^{\al}-
g^{\mu\al}x^{\nu})(f'(R_{0})R_{0}-f(R_{0}))\nom\\
& = & \pa_{\al}[f'(R_{0})\eta^{\mn\al}+\frac{1}{6\kappa^{2}}(g^{\mn}x^{\al}-
g^{\mu\al}x^{\nu})(f'(R_{0})R_{0}-f(R_{0}))]
\eea
so that the generalized Landau-Lifshitz superpotential takes the form
\beq
\tilde{h}^{\mn\al}=f'(R_{0})\eta^{\mn\al}+\frac{1}{6\kappa^{2}}(g^{\mn}x^{\al}-
g^{\mu\al}x^{\nu}) \left[f'(R_{0})R_{0}-f(R_{0})\right].
\eeq
The EMPT $t^{\mn}$ in a general coordinate system defined in the
Landau-Lifshitz prescription can now be read out from the expression for the EMC
(remember, that $g=-1$)
\be{kokokonti}
    \tau^{\mn}\equiv T^{\mn}+t^{\mn}\equiv \pa_{\al}\tilde{h}^{\mn\al}.
\ee
Hence the generalized Landau-Lifshitz EMC reads as
\beq\label{llemc}
\tau^{\mn}=f'(R_{0})\tau^{\mn}_{LL} +\inv{6\kappa^{2}}\left[f'(R_{0})R_{0}-f(R_{0})
\right]\pa_{\al}
\left(g^{\mn}x^{\al}-g^{\mu\al}x^{\nu}\right),
\eeq
where $\tau^{\mn}_{LL}$ is the Landau-Lifshitz EMC evaluated in the framework of GR
(see Eq. (\ref{kokokontiantti1})). The $00$-component reads as
\begin{widetext}
\bea{no16}
\tau^{00} & = & f'(R_{0})\tau^{00}_{LL}+\inv{6\kappa^{2}}
\left[f'(R_{0})R_{0}-f(R_{0})\right]
\pa_{\al}\left(g^{00}x^{\al}-g^{0\al}x^{0}\right)\nom\\
&=&f'(R_{0})\tau^{00}_{LL}+\frac{1}{6\kappa^{2}}(f'(R_{0})R_{0}-f(R_{0}))
\left(\pa_{i}g^{00}x^{i} + 3g^{00}\right).
\eea

Eq. (\ref{llemc}) is a general formula valid for any $f(R)$ theory when the studied metric
has constant scalar curvature. The standard GR result is recovered when $f(R)=R$.

\subsection{Energy-momentum complex of the SdS metric of some $f(R)$ models}

Using Eq. (\ref{llemc}) we can compute the EMC of the SdS metric in a general $f(R)$ theory:
\bea{no17}
\tau^{00}&=&f'(R_{0})\tau^{00}_{LL}+\frac{1}{6\kappa^{2}}(f'(R_{0})R_{0}-f(R_{0}))
\left(r B'(r) + 3B(r)\right)\nonumber\\
&=&f'(R_{0})\tau^{00}_{LL}+\frac{1}{6\kappa^{2}}(f'(R_{0})R_{0}-f(R_{0}))
\left(3-\frac{4 M}{r}- \frac{5\Lambda}{3}r^{2}\right)\nonumber\\
&=&-\frac{2}{\kap^2} \frac{-9\Lam r^4+(6M+\Lam r^3)^2}{r^2(\Lam r^3-3r+6M)^{2}}
f'(R_{0})+\frac{1}{6\kappa^{2}}(f'(R_{0})R_{0}-f(R_{0}))\left(3-\frac{4 M}{r}-
\frac{5\Lambda}{3}r^{2}\right).
\eea
\end{widetext}
This result is valid for any $f(R)$ theory that has the SdS metric as a vacuum solution \ie
any theory which satisfies the vacuum equation $f'(R_{0})R_{0}-2f(R_{0})=0$.
Again note that when $f(R)= R$ we recover the standard, \ie GR, form of Landau-Lifshitz EMPT,
Eq. (\ref{sdsll}), as expected.

An important special case encompassing popular choices of $f(R)$ is a generic action
function
\beq\label{darkinflaatio}
f(R)=R-(-1)^{n-1}\frac{a}{R^{n}}+(-1)^{m-1}b\, R^m,
\eeq
where $n$ and $m$ are positive integers and $a,b$ any real numbers. This form of function $f(R)$
is widely used in cosmological context.
In this case the generalized Landau-Lifshitz EMC takes the form
\begin{widetext}
\bea{pimeainfltau}
\tau^{00}&=&\frac{2^{-(1+2n)}\Lambda^{-n}}{9 r^{2}\kap^{2}}\biggl\{r\left[(12M-9r+5\Lambda r^{3})
(a(1+n)+b(m-1)(4\Lambda)^{m+n})\right]\nom\\
&-&9\frac{\left[-9\Lambda r^{4}+(6M+\Lambda r^{3})^{2}\right]\left[a\, n+(4\Lambda)^{n}
(4\Lambda+b\, m(4\Lambda)^{m})\right]}{\Lambda(6M-3r+\Lambda r^{3})^{2}}\biggr\}.
\eea
\end{widetext}
For the form of $f(R)$ considered above, i.e.
Eq. (\ref{darkinflaatio}), and recalling that for the SdS metric we have $R_0=-4\Lambda$,
constant curvature condition
can be written as
\beq
(4\Lambda)^{n+1}=a(n+2)+b(m-2)(4\Lambda)^{m+n}.
\eeq
In the special case where $m=2$ or $b=0$ we get
\be{bvakion}
a=\frac{(4\Lambda)^{n+1}}{n+2}.
\ee
Note, that not all type (\ref{darkinflaatio}) models are cosmologically viable. It is
known, that those vacuum solutions with $R_0$ such that $f''(R_0)>0$ (note our sign convention) are inherently
unstable \cite{dolgov} and therefore not suitable for cosmological model.

Particularly often used model of $f(R)$ theory of gravity is
\begin{equation}
f(R)=R-\frac{\mu^4}{R}-\eps R^{2},
\end{equation}
which has a stable vacuum whenever $\eps > 1/(3 \sqrt 3 \, \mu^2)$.
The $00$-component of the corresponding  generalized Landau-Lifshitz EMC for this model
is written as
\beq
\tau^{00}=\frac{1}{18 r^{2}\kap^{2}R_{0}}\biggl\{r(\eps R_{0}^{3}-2\mu^4)\left[12M-9r+5r^{3}
\Lambda\right]-36\frac{\left[-9\Lambda r^{4}+(6M+\Lambda r^{3})^{2}\right]\left[\mu^4+R_{0}^{2}-
2\eps R_{0}^{3}\right]}{R_{0}(6M-3r+\Lambda r^{3})^{2}}\biggr\},
\eeq
which for the special case of the SdS black hole metric with the cosmological
vacuum 
$R_0=-\sqrt{3}\, \mu^2$, reduces to
\beq
\tau^{00}= \frac{2+3\sqrt{3}\,\eps \mu^2}{18\sqrt{3} r^{2}\kap^{2}}\biggl\{\frac{5}{4}\sqrt{3}\mu^4 r^{4}
+3\mu^2 r(4M-3r)-72\frac{\left[192\sqrt{3}M^{2}+48\mu^2 M r^{3}-36\mu^2 r^{4}
+\sqrt{3}\mu^4r^{6}\right]}{(24M-12r+\sqrt{3}\mu^2 r^{3})^{2}}\biggr\}.
\eeq

Another cosmologically interesting $f(R)$ gravity model includes also logarithmic dependence on
curvature. Thus it reads
\beq
f(R)=R+(-1)^{m-1}c\, R^m-d\ln\left(\frac{|R|}{k}\right),
\ee
where its parameters are related to the cosmological constant by the constant curvature condition
written now as
\beq
d+(4\Lambda)^{m}c\, m=2\left[(4\Lambda)^{m}c +2\Lambda+d\ln\left(\frac{4\Lambda}{k}\right)\right].
\eeq
For this model the corresponding $00$-component of the Landau-Lifshitz EMC is of the form
\bea{no19}
\tau^{00}&=&\inv{18\kap^{2}r^{2}}\Big\{-9\frac{\left(d+c\,m(4\Lambda)^{m}+4\Lambda\right)
\left(-9\Lambda r^{4}+(6M+r^{3}\Lambda)^{2}\right)}{\Lambda\left(6M-3r+\Lambda r^{3}\right)^2}\nom\\
&+& r\left(12M-9r+5r^{3}\Lambda\right)\left(d+c(m-1)(4\Lambda)^{m}-d\ln\left(\frac{4\Lambda}{k}\right)
\right)\Big\}.
\eea

In any case the generalized Landau-Lifshitz EMPT of a $f(R)$ model differs crucially from the GR
Landau-Lifshitz EMPT for the SdS metric. Taking into account the constant curvature condition
we can write
\beq\label{gentau}
\tau^{00}= f'(R_{0})\tau^{00}_{LL}+\frac{1}{6\kappa^{2}}f(R_{0})
\left(3-\frac{4 M}{r}- \frac{5\Lambda}{3}r^{2}\right),
\eeq
which coincides with the GR Landau-Lifshitz EMPT only if $f(R_{0})=0$ and $f'(R_{0})=1$
(implying, due to the constant curvature condition, that there is no cosmological constant). 
This special case, while possible,
is not a general property of physically meaningful $f(R)$ models indicating that in a general $f(R)$ model
the Landau-Lifshitz EMPT will be non-trivially related to the corresponding EMPT in GR. For example,
it is clear that at large $r$ the two EMPTs have different asymptotic limits with $\tau_{LL}\sim r^{-2}$
in GR and $\tau_{LL}\sim r^{2}$ in a general $f(R)$ model.

\section{Conclusions and Discussion}

The problem of energy localization has been one of the first problems that was treated after the onset
of GR. Although a number of scientists endeavored to solve it, the energy localization remains a vexed
and unsolved problem till to date.
In this work, motivated by the recent interest in constructing extended models of gravity and in
particular $f(R)$ gravity models that replace the standard Einstein-Hilbert action of GR, we have introduced
for the very first time, to our knowledge, the energy localization problem in the framework of $f(R)$
theories of gravity. In particular, we have extended the concept of energy-momentum complex in the
prescription of Landau-Lifshitz. Although
we are unable to formulate a completely general expression for the
EMC valid for all theories and metrics, we can proceed in an important special case where the scalar
curvature of the considered metric is constant. In this case, we have presented a general formula for the
Landau-Lifshitz energy-momentum complex for a general $f(R)$ theory. We find that the general relativity
result is generalized to encompass an additional term.

Metrics satisfying the requirement of constant scalar curvature
include the Schwarzschild-de Sitter metric, which \eg describes the space-time around spherically symmetric
objects in a universe with a cosmological constant.
We have computed the generalized Landau-Lifshitz EMC for a general $f(R)$ theory that accepts the SdS metric
as a solution as wells as for a number of $f(R)$ commonly considered in the literature.
We find that the GR result is generalized by the presence of additional term. The new term is non-trivial
as it has a different dependence on the coordinate $r$ than the term arising from the GR part.

It is more than obvious that further study is needed, {\it e.g.} other EMC's and their interpretation, {\it i.e.}
corresponding physical boundary conditions,
in $f(R)$ models need to be considered. A particularly interesting and a potentially fruitful direction
to follow in the future is to consider the problem of energy localization in Weinberg's formulation.
For a more general EMC covering also the non-constant curvature case, a construction of a new type of EMC may be
a more direct way to proceed as generalization of the Landau-Lifshitz EMC is challenging.
The calculations of the integrated constants of motion in different models and systems is another example
of a relevant open question. We hope to address these issues in future work.

\acknowledgments
TM is supported by the Academy of Finland. AP acknowledges support from the Academy of Finland under project no.
8111953.



\begin{thebibliography}{X}

\bibitem{einstein} A. Einstein, Preuss. Akad. Wiss. Berlin
{\bf47}, 778 (1915); Addendum-ibid. {\bf47}, 799 (1915).

\bibitem{ll} L.D. Landau and E.M. Lifshitz, {\it The Classical Theory of
Fields}, (Addison-Wesley Press, Reading, MA), p. 317 (1951).

\bibitem{tolman} R.C. Tolman, {\it Relativity, Thermodynamics and
Cosmology}, (Oxford University Press, London), p. 227 (1934).

\bibitem{pp} A. Papapetrou, Proc. R. Ir. Acad. A {\bf52}, 11 (1948).



\bibitem{berg} P.G. Bergmann and R. Thompson, Phys. Rev. {\bf89},
400 (1953).

\bibitem{gold} J.N. Goldberg, Phys. Rev. {\bf111}, 315 (1958).

\bibitem{weinberg} S. Weinberg, {\it Gravitation and Cosmology: Principles
and Applications of General Theory of Relativity}, (Wiley, New
York), p. 165 (1972).

\bibitem{moller} C. M{\o}ller, Ann. Phys. (N.Y.) {\bf4}, 347
(1958).

\bibitem{drawbacks1}
S. Chandrasekhar and V. Ferrari, Proc. R. Soc. London A {\bf435}, 645 (1991).

\bibitem{drawbacks2}
G. Bergqvist, Class. Quant. Grav. {\bf9}, 1753 (1992).

\bibitem{drawbacks3}
G. Bergqvist, Class. Quant. Grav. {\bf9}, 1917 (1992).

\bibitem{drawbacks4}
C.M. Chen and J.M. Nester, Class. Quant. Grav. {\bf16}, 1279 (1999).

\bibitem{chandra} S. Chandrasekhar and V. Ferrari, Proc. R. Soc.
London A {\bf435}, 645 (1991).

\bibitem{bergqvist1} G. Bergqvist, Class. Quant. Grav. {\bf9}, 1753
(1992).

\bibitem{bergqvist2} G. Bergqvist, Class. Quant. Grav. {\bf9},
1917 (1992).

\bibitem{chiang} C.M. Chen and J.M. Nester, Class. Quant. Grav.
{\bf16}, 1279 (1999).








\bibitem{virbhadra1}
K.S. Virbhadra, Phys. Rev. D {\bf41}, 1086 (1990).

\bibitem{virbhadra2}
K.S. Virbhadra, Phys. Rev. D {\bf42}, 1066 (1990).

\bibitem{virbhadra3}
K.S. Virbhadra, Phys. Rev. D {\bf42}, 2919 (1990).

\bibitem{virbhadra4}
K.S. Virbhadra, Pramana {\bf38}, 31 (1992).

\bibitem{virbhadra5}
N. Rosen and K.S. Virbhadra, Gen. Rel. Grav. {\bf25}, 429 (1993).

\bibitem{virbhadra6}
K.S. Virbhadra and J.C. Parikh, Phys. Lett. B {\bf317}, 312 (1993).

\bibitem{virbhadra7}
K.S. Virbhadra and J.C. Parikh, Phys. Lett. B {\bf331}, 302 (1994); Erratum-ibid B {\bf340}, 265 (1994).

\bibitem{virbhadra8}
K.S. Virbhadra, Pramana {\bf44}, 317 (1995).

\bibitem{virbhadra9}
K.S. Virbhadra, Pramana {\bf45}, 215
(1995).



\bibitem{virbhadra11}
A. Chamorro and K.S. Virbhadra, Int.J. Mod. Phys. D {\bf5}, 251 (1997).

\bibitem{Vagenas:2006pj}
  E.~C.~Vagenas,
  Mod.\ Phys.\ Lett.\  A {\bf 21}, 1947 (2006)
  [arXiv:gr-qc/0602107].

\bibitem{virbhadra10}
J.M. Aguirregabiria, A.Chamorro and K.S. Virbhadra, Gen.
Rel. Grav. {\bf28}, 1393 (1996).

\bibitem{pen} R. Penrose, Proc. R. Soc. London A {\bf381}, 53
(1982).

\bibitem{tod} K.P. Tod, Proc. R. Soc. London A {\bf388}, 457
(1983).




\bibitem{mann1} J.D. Brown, J. Creighton, and R.B. Mann, Phys. Rev. D {\bf50},
6394 (1994).

\bibitem{quasilocal1} S.A. Hayward, Phys. Rev. D {\bf49}, 831 (1994).

\bibitem{quasilocal2}
S.W. Hawking and G.T. Horowitz, Class. Quant. Grav. {\bf13},  1487 (1996).

\bibitem{quasilocal3}
J.D. Brown, S.R. Lau, and J.W. York, Phys. Rev. D {\bf55}, 1977 (1997).

\bibitem{quasilocal4}
S.-T. Yau, Adv. Theor. Math. Phys. {\bf5}, 755 (2001).

\bibitem{quasilocal5}
C.-C.M. Liu and S.-T. Yau, Phys. Rev. Lett. {\bf90}, 231102 (2003).


\bibitem{senovilla1} J.M.M. Senovilla, Class. Quant. Grav. {\bf 17}, 2799 (2000).


\bibitem{bel1}
L. Bel, C.R. Acad. Sci. Paris {\bf247}, 1094 (1958).

\bibitem{bel2}
L. Bel, PhD Thesis (CDU et SEDES Paris 5e) (1960).

\bibitem{bel3}
L. Bel, Commun. Math. Phys. {\bf138}, 59 (1962).



\bibitem{robinson}I. Robinson, Class. Quant. Grav. {\bf14}, A331 (1997).

\bibitem{senovilla2}
M.A.G. Bonilla and J.M.M. Senovilla, Phys. Rev. Lett. {\bf78}, 783 (1997).

\bibitem{senovilla3}
M.A.G. Bonilla and J.M.M. Senovilla, Gen. Rel. Grav. {\bf29}, 91 (1997).

\bibitem{senovilla4}
J.M.M. Senovilla, {\it Remarks on superenergy tensors}, to appear in ``Gravitation and Relativity in General'', (World Scientific, 1999),
gr-qc/9901019 .

\bibitem{senovilla5}
P. Teyssandier, {\it Superenergy tensors for a massive scalar field}, gr-qc/9905080; J. Garecki,
Annalen Phys. {\bf10}, 911 (2001).

\bibitem{senovilla6}
R. Lazkog, J.M.M. Senovilla, and R. Vera, Class. Quant. Grav. {\bf 20}, 4135 (2003).

\bibitem{nester} C.C. Chang, J.M. Nester, and C.M. Chen, Phys. Rev.
Lett. {\bf83}, 1897 (1999).





\bibitem{turner}  S.~M.~Carroll, V.~Duvvuri, M.~Trodden and M.~S.~Turner,
Phys.\ Rev.\ D {\bf 70}, 043528 (2004).

\bibitem{turner2}  S.~M.~Carroll, A.~De Felice, V.~Duvvuri, D.~A.~Easson, M.~Trodden
and M.~S.~Turner, Phys.\ Rev.\ D {\bf 71}, 063513 (2005).

\bibitem{allemandi}  G.~Allemandi, A.~Borowiec and M.~Francaviglia,
Phys.\ Rev.\ D {\bf 70}, 103503 (2004).

\bibitem{meng}  X.~Meng and P.~Wang,
Class.\ Quant.\ Grav.\  {\bf 21}, 951 (2004).

\bibitem{nojiri3}  S.~Nojiri and S.~D.~Odintsov,
Phys.\ Rev.\ D {\bf 68}, 123512 (2003).

\bibitem{cappo1} S.~Capozziello, Int.\ J.\ Mod.\ Phys.\ D {\bf 11},
483 (2002).

\bibitem{nojiri2}  S.~Nojiri and S.~D.~Odintsov,
Phys.\ Lett.\ B {\bf 576}, 5 (2003).

 \bibitem{woodard}
  R.~P.~Woodard,
  Lect.\ Notes Phys.\  {\bf 720}, 403 (2007)
  [arXiv:astro-ph/0601672].
\bibitem{odintsov} For discussion of other generalized theories, see {\it e.g.}:
 S.~Nojiri and S.~D.~Odintsov, arXiv:hep-th/0601213.

\bibitem{new2}
  S.~Tsujikawa,
  Phys.\ Rev.\  D {\bf 77}, 023507 (2008)
  [arXiv:0709.1391 [astro-ph]].


\bibitem{Nojiri:2007as}
  S.~Nojiri and S.~D.~Odintsov,
  Phys.\ Lett.\  B {\bf 657}, 238 (2007)
  [arXiv:0707.1941 [hep-th]].

\bibitem{Starobinsky:2007hu}
  A.~A.~Starobinsky,
  JETP Lett.\  {\bf 86}, 157 (2007)
  [arXiv:0706.2041 [astro-ph]].

\bibitem{dolgov}  A.~D.~Dolgov and M.~Kawasaki,
Phys.\ Lett.\ B {\bf 573}, 1 (2003).

\bibitem{soussa}  M.~E.~Soussa and R.~P.~Woodard,
 Gen.\ Rel.\ Grav.\  {\bf 36}, 855 (2004)  [arXiv:astro-ph/0308114].

\bibitem{faraoni}   V.~Faraoni and S.~Nadeau,
Phys.\ Rev.\ D {\bf 72}, 124005 (2005)  [arXiv:gr-qc/0511094].

\bibitem{chiba}  T.~Chiba,
Phys.\ Lett.\ B {\bf 575}, 1 (2003).

\bibitem{confprobs}  E.~E.~Flanagan, Class.\ Quant.\ Grav.\  {\bf 21}, 417 (2003);
Class.\ Quant.\ Grav.\  {\bf 21}, 3817 (2004);
G.~Magnano and L.~M.~Sokolowski,
Phys.\ Rev.\ D {\bf 50}, 5039 (1994).

\bibitem{Clifton}  T.~Clifton and J.~D.~Barrow,
 Phys.\ Rev.\ D {\bf 72}, 103005 (2005)  [arXiv:gr-qc/0509059].

\bibitem{Hu2007}
  W.~Hu and I.~Sawicki,
  Phys.\ Rev.\  D {\bf 76}, 064004 (2007)
  [arXiv:0705.1158 [astro-ph]].

\bibitem{Bean:2006up}
  R.~Bean, D.~Bernat, L.~Pogosian, A.~Silvestri and M.~Trodden,
  Phys.\ Rev.\  D {\bf 75}, 064020 (2007)
  [arXiv:astro-ph/0611321].


\bibitem{Song:2006ej}
Y.~S.~Song, W.~Hu and I.~Sawicki,
  Phys.\ Rev.\  D {\bf 75}, 044004 (2007)
  [arXiv:astro-ph/0610532].

\bibitem{Song2}
  Y.~S.~Song, H.~Peiris and W.~Hu,
  Phys.\ Rev.\  D {\bf 76}, 063517 (2007)
  [arXiv:0706.2399 [astro-ph]].

\bibitem{Erickcek:2006vf}
A.~L.~Erickcek, T.~L.~Smith and M.~Kamionkowski,
  Phys.\ Rev.\  D {\bf 74}, 121501 (2006)
  [arXiv:astro-ph/0610483].

\bibitem{Chiba2}
 T.~Chiba, T.~L.~Smith and A.~L.~Erickcek,
  Phys.\ Rev.\  D {\bf 75}, 124014 (2007)
  [arXiv:astro-ph/0611867].

\bibitem{Jin:2006if}
X.~H.~Jin, D.~J.~Liu and X.~Z.~Li,
arXiv:astro-ph/0610854.  

\bibitem{Faulkner:2006ub}
  T.~Faulkner, M.~Tegmark, E.~F.~Bunn and Y.~Mao,
  Phys.\ Rev.\  D {\bf 76}, 063505 (2007)
  [arXiv:astro-ph/0612569].

\bibitem{kimmo}
  K.~Kainulainen, J.~Piilonen, V.~Reijonen and D.~Sunhede,
  Phys.\ Rev.\  D {\bf 76}, 024020 (2007)
  [arXiv:0704.2729 [gr-qc]].

\bibitem{Nojiri:2007cq}
  S.~Nojiri and S.~D.~Odintsov,
  arXiv:0710.1738 [hep-th].

\bibitem{Cognola:2007zu}
  G.~Cognola, E.~Elizalde, S.~Nojiri, S.~D.~Odintsov, L.~Sebastiani and S.~Zerbini,
  arXiv:0712.4017 [hep-th].

\bibitem{Cognola:2005de}
  G.~Cognola, E.~Elizalde, S.~Nojiri, S.~D.~Odintsov and S.~Zerbini,
  JCAP {\bf 0502}, 010 (2005)
  [arXiv:hep-th/0501096].

\bibitem{Multamaki2}  T.~Multamaki and I.~Vilja,
Phys.\ Rev.\ D {\bf 74}, 064022 (2006)  [arXiv:astro-ph/0606373].

\bibitem{virbhadra12} K.S. Virbhadra, Phys. Rev. D {\bf60}, 104041 (1999).

\bibitem{xuludoctor}
  S.~S.~Xulu,
  arXiv:hep-th/0308070.

\bibitem{bak}
  D.~Bak, D.~Cangemi and R.~Jackiw,
  Phys.\ Rev.\  D {\bf 49}, 5173 (1994)
  [Erratum-ibid.\  D {\bf 52}, 3753 (1995)]
  [arXiv:hep-th/9310025].

\bibitem{koivisto}
  T.~Koivisto,
  Class.\ Quant.\ Grav.\  {\bf 23}, 4289 (2006)
  [arXiv:gr-qc/0505128].

\end{thebibliography}
\end{document}